# Lattice Percolation Approach to Numerical Modeling of Tissue Aging

Vladimir Privman,[a]  Vyacheslav Gorshkov,[b]  and  Sergiy Libert[c,]*

[a]Department of Physics, Clarkson University, Potsdam, NY 13699, USA
[b]National Technical University of Ukraine — KPI, Kiev 03056, Ukraine
[c]Department of Biomedical Sciences, Cornell University, Ithaca, NY 14853, USA

## Abstract

We describe a percolation-type approach to modeling of the processes of aging and certain other properties of tissues analyzed as systems consisting of interacting cells. Tissues are considered as structures made of regular healthy, senescent, dead (apoptotic) cells, and studied dynamically, with the ongoing processes including regular cell division to fill vacant sites left by dead cells, healthy cells becoming senescent or dying, and other processes. Statistical-mechanics description can provide patterns of time dependence and snapshots of morphological system properties. An illustrative application of the developed theoretical modeling approach is reported, confirming recent experimental findings that inhibition of senescence can lead to extended lifespan.

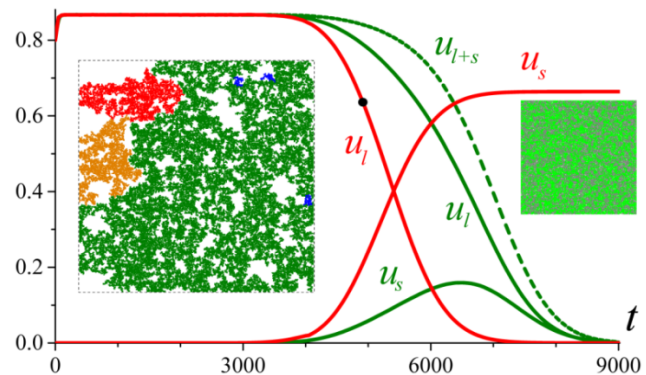

## Keywords

Tissue aging, percolation model, senescence

---

*E-mail: libert@cornell.edu

# Introduction

Aging of organisms or tissues can be viewed as a statistical-mechanics problem of the loss of homeostasis (approximate steady state) and integrity over time. Processes that ensure tissue homeostasis and integrity, such as healing after an injury, are a topic of a major interest [1, 2] to both medical professionals from an application point of view [3] and to basic scientists from a mechanistic point of view [4]. In this work we consider potential applications of lattice percolation-type modeling to develop a quantitative description of the processes of aging of tissues analyzed as structures consisting of interacting cells. Percolation models provide information and predictions on the system's integrity and connectivity [5, 6]. Here we argue that, tissues can be considered as structures made of regular healthy cells, senescent cells, dead (apoptotic) cells, and can be studied dynamically, with the ongoing processes involving regular cells' division to fill "vacant sites" left by dead cells, regular cells becoming senescent or dying, senescent cells disrupting functioning of the nearby regular cells, etc. The obtained connectivity information for various cell types can then be related to the physical or biological tissue properties and structure.

Such a statistical-mechanics description can capture interesting patterns of time dependence, especially if we want to consider the possible effects of active interventions, such as application of certain drugs on the ongoing cell dynamics in healthy or diseased (for example cancerous) tissue. This expectation is suggested by recent studies [7-10] that have applied percolation ideas to self-healing and self-damaging in devices and materials — an approach that in itself can be regarded as a utilization of bio-inspired concepts. However, application of percolation-modeling ideas to the dynamics of aging in tissues, rather than to the onset of the initial-fatigue damage in materials requires new connectivity and dynamical rules, different ideas and interpretations, and, ultimately, relating the modeling approach to a different set of experimental observations. Regarding the latter, there are much more experimental data presently available for damages in tissues [11-14], especially when it is locally caused (for example, skin biopsy), than for the micro- or nano-scale autonomous (locally triggered) self-damaging [15, 16] in materials. Note that much more data are available on autonomous self-healing in materials, when damage buildup is avoided or delayed by incorporating active capsules [17-33].

There is an increasing interest in developing "smart materials" that have advanced properties, such as autonomous self-healing. To model these materials, earlier studies of percolation models that involve various "cells" on a lattice, with the "cells" interacting with each other resulting in the change in their cluster structure and therefore connectivity have been reported [7-9, 17-22, 34, 35]. In such materials, development of damage and fatigue can be delayed by embedded capsules containing a microcrack-healing agent activated by a local "triggering" mechanism [9]. Very recently there has also been interest in materials with autonomous self-damaging properties [10, 15, 16], as well as in situations when both mechanisms are utilized [10]. The first autonomous self-healing of polymer composites was realized [19] with the polymerization process initiated by the healing agent (released from cracked capsules) preventing the propagation of cracks caused by mechanical stress. This finding was followed by interesting experimental [7, 18, 20, 21, 23-33] and theoretical [7-10, 34-40] developments. A promising new application of smart-materials concepts could be in the development of "self-damaging" systems, devices, and materials, also termed self-destructive, self-deteriorating, or transient. The first experimental realization of this property at the device-component level aimed at injectable devices has recently been reported [15, 16] for the "transient electronics." Such bio-inspired concepts can be beneficial for many applications.

One important avenue of research has involved achieving "smart" response at the nanoscale [7-9, 21, 22, 41]. The reason for this has been that such material designs promise to allow autonomous control of the material's fatigue at the earliest stages of its development, when damage is not yet macroscopic. Material degradation [42] ultimately results in formation of voids and cracks on the macroscopic level. These are initiated by the development of microscopic crazes and microcracks, the growth of which can be prevented (for self-healing) or accelerated (for self-damaging) at the nanoscale.

Here we devise a percolation approach to model dynamics of live tissues and connect the results to biological observations. The modeling is rooted in approaches used for materials [7-10], however, there are obvious and significant differences. Indeed, in tissues each cell is an "active element," not only changing in time but also affecting other cells. Challenges in model development will include connecting such approaches to quantitative biological experimental data and applied the modeling results to better understand aging. We will also mention the possible extensions of our modeling setup to wound healing and tissue engineering. We consider a percolation modeling approach to study probabilistic changes in the cell occupancy in a lattice model, offering a microscopic statistical-mechanics description of the time-dependent tissue connectivity. This requires consideration of various types of cells, such as regular "live" cells, dead cells and vacant sites left after cells are eliminated, and introduction of biological processes, such as cell division, which can heal damage, cell senescence (permanent withdrawal of the cell from cycling, i.e., division), and apoptosis (cell self-destruction in response internal irreparable damage or infection). The time-evolution of the cell occupancy at lattice sites can then be followed numerically based on a set of rules that will have to be defined based on existing biological data, and is described in Methods section.

Statistical mechanics is a broad field of research and application, many techniques and concepts of which have been applied to biological or bio-inspired systems. We are not aware of studies of the type reported here, of cellular interactions in the context of them resulting in tissue aging and related time-dependent effects. We point out, however, that statistical-mechanics approaches to aging processes *within* cells and their relation to cancer — avoidance of which is generally intimately connected with processes that also control aging [43] — have been recently reported [44, 45]. On the cellular scale, our lattice-percolation approach is a complicated version of cellular automata, which is a field of study of pattern emergence. Certain other applications of percolation modeling to biological problems have been reported [46, 47]. Most statistical-mechanics studies, however, have been focused on scaling at critical points or properties of large clusters, which are not regimes of primary interest in the present context.

In the Methods section, we motivate and introduce our theoretical modeling approach. We then offer an illustrative application, as described in the Results section, which yields results matching recent experimental findings that inhibiting senescence can extend lifespan [48, 49]. We will follow with the Discussion section to generalize about how modeling results can be related to pre-designed experiments tailored to probe cell "connectivity" properties. We point out, however, that, similar to the studies of self-healing/self-damaging in materials, here we do not consider the critical-point behavior regime near the percolation transition, which has been of interest in many studies [5, 6] of percolation models, because we are only interested in the regime of relatively high connectivity, especially for regular cells (before the tissue loses its cohesion/identity, literally, "dies"). Additionally, in the Discussion, we will address heterogeneous dynamics involving focal external damage (such as cuts), and other effects that can be considered, including weak systemic damage (such as exposure to radiation), external delivery of different types of cell to large wounded areas to facilitate healing, which is reminiscent of stem cell therapy, and cell dynamics in the presence of scaffolds (tissues engineering).

**Methods**

*Utility of lattice models in studying tissue aging and other biological properties*

Microscopic statistical-mechanics modeling constitutes one of the possible approaches to explore patterns of possible behaviors involving various rules of cell interactions. The degree of complexity and local correlations that are required for percolation-type models to mimic observed tissue aging, healing, and other properties are of interest. The actual macroscopic physical properties are typically nonlinear in the microscopic morphology (connectivity) measures. For materials, electrical transport properties are among the most natural to calculate in the percolation-model context [7-10], by connecting sites with bonds of conductance that depends on the nature of the sites' occupancy. The conductivity properties can also be measured for tissues. Elastic properties of the system can be explored in the percolation-model context [50-53], by decorating bonds with springs, the rigidity of which depends on the site occupancy. Furthermore, in the Discussion section we outline possible experiments that probe various cell-type statistics and also cell-clustering morphological properties, rather than mechanical or conductance properties.

Generally, on the microscopic scale, statistical-mechanics descriptions of the cell behavior within the tissue can be "cartoonish" in nature, capable of encompassing only general features of the actual microscopic morphology. However, such information is of interest, since the description of the time-dependence of various properties of regions and systems much larger than a few cells will help us understand how to control the system's complex processes to achieve desired properties (such as accelerated healing, or resistance to damage). Within such an approach, concepts of programmed active local healing and damaging have been of interest not only in the studies of materials, but also in designs of sensor networks' functionality [54-56], especially when active response is desired: Self-damaging concepts can be useful, for example, in the designs aimed at abruptly shutting a whole sensor network down if enough interconnected nodes have become compromised.

Considering that there is no critical-point universality for percolation at high connectivity, which is a regime of interest here, one can legitimately ask what specific quantitative and qualitative results can be expected from lattice-model studies for tissues, as well as earlier studies for materials. On the level of phenomenological data fitting, models of the type considered in the following subsection can yield expressions for a few-parameter fitting of data for measured tissue properties. These parameters are usually various process rates, which can thus be extracted and classified based on macroscopic measurements. On a more qualitative level, percolation and similar lattice models can offer insights into the degree of complexity required to achieve specific time-dependent responses.

For example, for self-healing materials it was noted that relatively short-range healing (e.g., nearest neighbor) sufficed [8, 9] for useful effects, whereas a recent study of self-damaging indicated [10] that a much longer range (at least half a dozen neighbors away) action by damage-causing "capsules," and only when also combined with self-healing in complicated rules, was required to have the "abrupt shutdown" property described above. It is noteworthy that this correlates with the property that senescent cells cause damage in large neighborhoods [57], promising interesting findings regarding their role in tissue aging. The behaviors for some rules might not even be entirely monotonic. Some of these features are illustrated in Fig. 1.

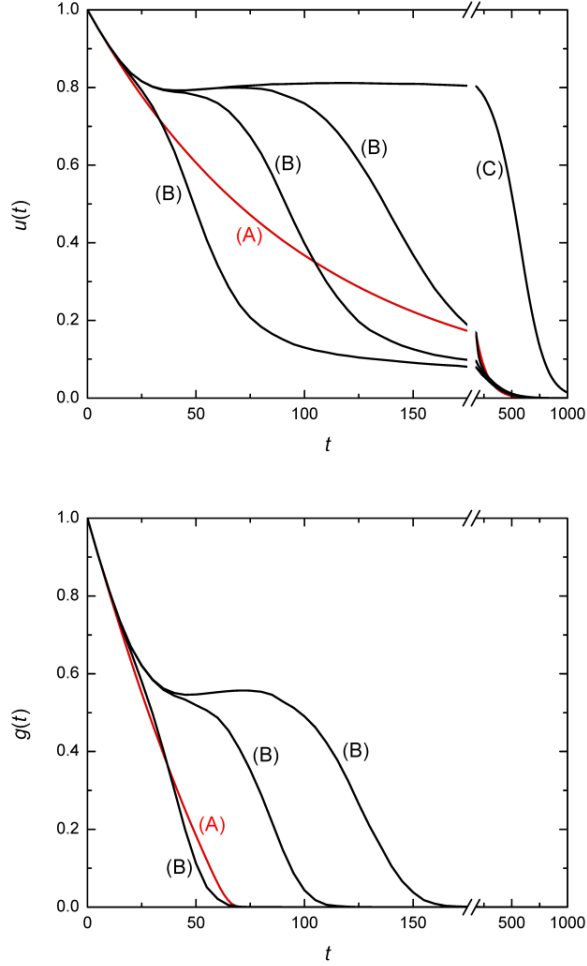

**Figure 1.** An example, following Ref. [10], for a model of a two-dimensional bond-percolation lattice, for the action of damaging sites, conditioned on the activation of healing sites. The top panel shows the measure of the microscopic "integrity:" the fraction of the bonds, $u(t)$, that are "intact" at time $t$. The bottom panel shows the conductivity of the lattice, $g(t)$, normalized per that of the fully intact lattice. The curves show: (A) no active damaging or healing by added capsules, only the natural "wear-and-tear" material degradation (here, bond breakage), corresponding to exponential decay of $u(t)$; (B) three examples of a different balance of added active damaging vs. active healing, with specific rules detailed in Ref. [10]; (C) the active healing process only, with no active damaging (the conductivity for this case is not shown in the bottom panel because $g(t)$ follows $u(t)$ approximately linearly for times up to $t \approx 200$).

*Homogeneous-system lattice percolation modeling of tissue aging*

The modeling approach can consider 2D (two-dimensional) tissue samples, for instance, layers such as in skin, as described in the Discusison section, that addresses a possible experimental approach, or 3D (three-dimensional) samples, idealized in both cases as lattices of sites, for example, the square (2D) or cubic (3D) lattices. We comment on this lattice-model idealization and lattice selection later (in the Discussion). Various cell (lattice site) populations can be considered, represented by sites of various "occupancy," such as regular healthy live cells, senescent cells, dead cells, vacant sites, etc. We use the Monte Carlo (MC) approach to the cells' dynamics. MC sweeps through the system are carried out according to probabilistic rules that attempt to mimic in the site-lattice description the features expected of the cells' influencing their own and nearby sites' occupancy. Typically, sites are probed at random, such that on average the full lattice will be swept over once, during each unit time step. This defines MC time, $t$, units, such as shown in Fig. 1 for a different problem (for a model of self-damaging/self-healing material). Averages over multiple runs are done for estimating the cluster-structure and connectivity statistics. Single-run snapshots can provide cluster morphology information (illustrated in the Results section).

Different organs and tissues of a human body employ slightly different strategies to ensure cellular homeostasis (approximate steady state of cell-type populations) and integrity. Let us outline a possible set of stochastic rules to define for numerical simulations, based on information of cell behavior, considering a 2D "skin-cell layer" case (see the experimental example in the Discussion section) as a convenient idealization to begin with. In addition to its relative simplicity, skin is also the organ that is the most susceptible to the extrinsic insults. Therefore, the strategies that naturally evolved to ensure the integrity and longevity of this organ are of utmost interest.

The following set of rules for the dynamic modeling is a simplified reflection on the observations made for the processes in aging and/or injured skin, more specifically, epidermis. The basal layer of epidermis is called *stratum basale*. It consists of the monolayer of keratynocytes, which divide when necessary to populate the upper layers of the skin, which serve as the physical barrier between the environment and the organism. The rules provide a lattice-model "cartoon" description of the cell dynamics resulting from the processes in this layer:

**1.** Regular cells can divide (cycle) to produce two identical cells as the result. The number of divisions the cell can undertake is, however, limited. The nature of this limit lies in the inability of the cell to fully replicate the ends of the chromosomes — telomeres, which at some point shorten to the limit at which cell senescence is triggered [58, 59]. In a lattice model a site with a "regular" cell, once picked during a MC-step sweep, will be assigned small probability to divide, as long as the count of its remaining "lifetimes" is not exceeded, or otherwise it can become senescent: see the next rule. In the former case both cells inherit the remaining "lifetimes" count (decreased by 1) and the newborn cell will fill an empty nearest-neighbor site if one is available. Otherwise, the division might not be carried out at all if contact inhibition prevents cell division in a locally fully packed region, or the "daughter" cell can be discarded provided we consider the situation in a single 2D layer, or advance to another layer following more complicated rules, if we go beyond 2D simulation, such as modelling of scar formation, etc.

**2.** Cells can senesce. Senescent cells are still alive, but they permanently withdraw from the cell cycling. Senescence of the cell can be triggered by either exhaustion of its division potential or by damage [55, 56]. The latter process can be modeled by assigning small probabilities (small rates as compared to the MC time step unit) for regular sites to undergo damage-related events: see the next rule.

**3.** Regarding cells being damaged: The source of the damage could be extrinsic (external physical or chemical insult) or intrinsic (inflammation or DNA mutation). Severely damaged cells will die instantly, such as physically raptured cells. Moderately damaged cells can either enter growth arrest (temporary senescence, to take time to repair possible DNA damage), permanent senescence, or apoptosis (programmed cell death), depending on the intensity of the injury. We can assume homogeneous (over the system) damage processes and/or heterogeneous modeling, the latter commented on in results and discussions. The probabilities (rates) of various processes vary locally for each randomly probed (during MC sweeps) cell depending not only on its own state but also on its neighborhood: see the next rule.

**4.** Dynamical effects need not be short-range at the probed sites or their closest neighbors. External physical or chemical insults turn on an inflammation response. Cells within the inflamed area are more susceptible to apoptosis. The role of the inflammation is to facilitate the pathogen clearance, but it is damaging to the host cells as well [60], the latter effect being caused by the secretion of factors, such as tumor necrosis factor-alpha (TNF-α), see Ref. [61], that affect various-size neighborhoods. Inflammatory dermal lesions (dermonecrosis induced by the immune system) after a local infection in animal studies reach approximately $100\,\text{mm}^2$, see Ref. [62]. Considering the average size of the keratinocyte [63] of about $5\times10^{-5}\,\text{m}$, we can estimate that the local inflammation signals in skin spread quite far: approximately across 110 cells in each direction. For homogeneous modeling, we can incorporate the rate of damage mentioned in Rule 2 to be partially attributed to the intrinsic damage and partially to the extrinsic events. Such inflammation events will affect various process rates for some time intervals in the extended neighborhood of the affected cells. Thus, some event rates will have "normal" and "inflamed" values, the latter reverting to normal (if no new nearby inflammation occurs) after some time, by decaying or increasing (see the next two rules) with some assumed time dependence.

**5.** Cells can undergo apoptosis [64, 65], which is a programmed cell death. Apoptosis can be triggered by irreparable DNA damage [66, 67] or severe insult. Cells are usually sensitized to apoptosis by inflammation.

**6.** Senescent cells are resistant to apoptosis [68]. This can be modelled by such cells "dying" at a significantly reduced rate.

**7.** There are in fact many possible sources of the probabilities of various processes not only influencing other future processes, but also triggering secondary events at the same site and/or its neighborhood with some probability of their own. Such processes should then also be incorporated in the MC simulation steps. Specifically, the process of the cell division is associated with the possibility of introducing DNA mutations and thus, damage [2]. Severe DNA damage introduced as the result of DNA replication during cell division can lead to premature senescence or apoptosis; cells are especially sensitive to stress (damage) during division.

**8.** Some processes can influence rates of future processes in relatively large neighborhoods. Senescent cells secrete factors that instruct surrounding cells to undergo cell division more rapidly, which increases the chances of the surrounding cells to accumulate DNA damage [69]. Factors secreted by senescent cells are known to induce a state of light chronic inflammation and thus also sensitize surrounding cells to apoptosis [57], which is another cell-cell "interaction" effect.

The above observations suggest that rather detailed models will have to be considered for stochastic dynamics to provide an accurate "cartoon" description of the cell populations as a function of time. The initial conditions, at time $t = 0$, for the modeling could be set as follows: Lattice filled very densely, to have a connected cluster of healthy live cells, well above the percolation threshold, with division limit set at 50, Ref. [70], which would represent healthy newborn skin tissue. For modeling the impact of extrinsic damage caused by environmental factors, such as radiation or chemotherapy on the adult human skin, the initial conditions ($t = 0$) can be set differently. For instance, have the division limit normally distributed for different cells, with an average value of 30 divisions left, and senescent cells randomly dispersed throughout the tissue with the frequency of $10^{-4}$.

The expected outcomes of numerical modeling can be as follows:

**(i)** Lattice modeling can explore the degree of complexity of dynamical rules that yield autonomously viable tissues, with "viable" defined according to possible measures such as long-duration survival of a substantial, approximately steady (in the site-type fractions), and well- and continuously-connected (in the largest, so-called "percolating" cluster) population of regular cells as long as the external damage events are not too severe. It will be interesting to understand to what extent are the naturally evolved rules "minimal" to achieve durable tissues viability.

**(ii)** Statistical and morphological properties of various cell states can be quantified (for the former) and visualized (for the latter), for example, size (site count) distribution and spatial interrelations of various cell-type connected structures (clusters).

**(iii)** We can also attribute physical properties to cells and their contacts (bonds between nearest neighbor lattice sites) and then evaluate macroscopic properties by established computational approaches, such as the system's conductivity or elasticity, mentioned earlier, etc. Physical properties can then also be used as measures of the tissue's "health."

The next section offers and illustration of the potential of the percolation approach, based on a very simple subset of rules from the full set of Rules 1-8 defined above, to explain specific experimental observations.

**Results**

For a subset of rules suggested by the list of properties above (Items 1-8), we report an illustrative study of two models: one without senescence and another with senescence present in a very simplified form. In this respect it is fascinating that removal of senescent cells in mice was found [48] to extend their longevity and ameliorate numerous age-associated pathologies: The senescence marker gene, $p16^{Ink4a}$, was genetically linked to activation of caspase 8, which is a trigger of apoptosis. Life-long removal of $p16^{Ink4a}$ expressing cells delayed acquisition of age-related pathologies in adipose tissue, skeletal muscle and eye. Interestingly, our results reported below, match this and other recent experimental findings [48, 49] that, inhibiting senescence extends lifespan.

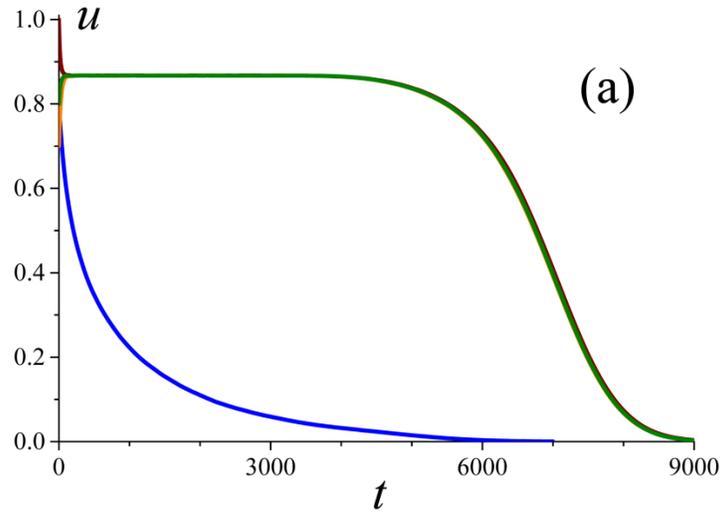

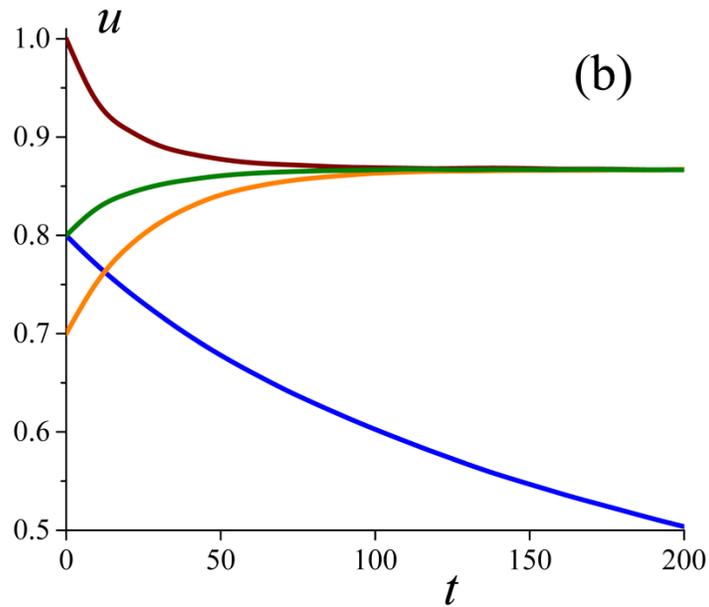

**Figure 2.** Model without senescence. **(a)** The bottom curve corresponds to a small rate of cell divisions (see text). The top three curves (largely overlapping) illustrate that, if the rate of cell divisions is large enough to achieve a proper balance with the rate of cell death (see text for details), and the geometrical restrictions are set as described in Rule 1, then a steady state is rapidly achieved and persists for extended times, though ultimately exponential decay takes over once cell division potential is exhausted. The plot shows such curves with the steady-state regime for three different initial healthy live cell densities. **(b)** Magnification of the initial short-time behavior.

Our MC simulations were for a square lattice of $500 \times 500$ sites, and all the plots reporting time-dependent, etc., quantities representing averages over at least 300 independent MC runs. Figure 2 shows results for a model without senescence. Cell division is limited to a total of 50 divisions, carried out only if one or more vacant nearest neighbors are available to be randomly selected for placing that "daughter" cell there. Once the cell division potential is exhausted, however, we do not, for the present model, consider them long-lived "senescent;" they can still die at the same rate as before. When the rate of cell division is properly balanced with cell death, a steady state is rapidly achieved. This steady state can persist for an extended time, Fig. 2, though ultimately exponential decay takes over once cell division potential is exhausted. A healthy live cell dies with probability $p \ll 1$ each time it is probed in a random MC sweep, or divides with probability $q \ll 1$ (as long as its count of possible divisions it not exceeded, in which case $q$ is instead set to zero), or remains unchanged with probability $1 - p - q$. Here we took $p = 0.010, q = 0.025$ (for the top three curves in Fig. 2). The vertical axis in Fig. 2 represents the fraction, $u$, of all the remaining cells (including those that can no longer divide), normalized per the total number of lattice sites; the horizontal axis represents the computational time (number of MC sweeps). We find that the steady state can persist for considerably longer time scales than the initial equilibration; this occurs when $q$ exceeds $p$ in the present model. However, if the division rate is set too low to be competitive with the death rate, then there is no steady state and the decay is practically exponential. Figure 2 illustrates this (the monotonically decaying bottom curve) for $p = 0.010, q = 0.010$. Without any divisions at all, the decay would be exactly exponential, as curve (A) in the top panel of Fig. 1.

Let us now introduce a simplified version of cell senescence (a variant of Rule 6). The above model is modified by designating all those cells that can no longer divide as long-lived senescent by assuming that, such cells are immortal on the time scales of all the other dynamical processes in the problem. For the original model (Fig. 2), we now separated, see Fig. 3, the counts of the healthy live cells that can divide, and those "senescent" cells that can no longer divide (even though in that model the latter cells still die at the same rate). Once the senescent cells are instead made long-lived (i.e., we not only set $q$ but also $p$ to zero for such cells), the cell counts (densities) for both cell types are modified, as shown in Fig. 3. Interestingly, the net effect of the senescent cells' introduction is that the lifespan of the "regular" healthy live cells is actually shortened: compare the two curves labelled $u_l$ in Fig. 3. This observation resonates with the recent experimental findings that inhibition of senescence can extend lifespan [48, 49].

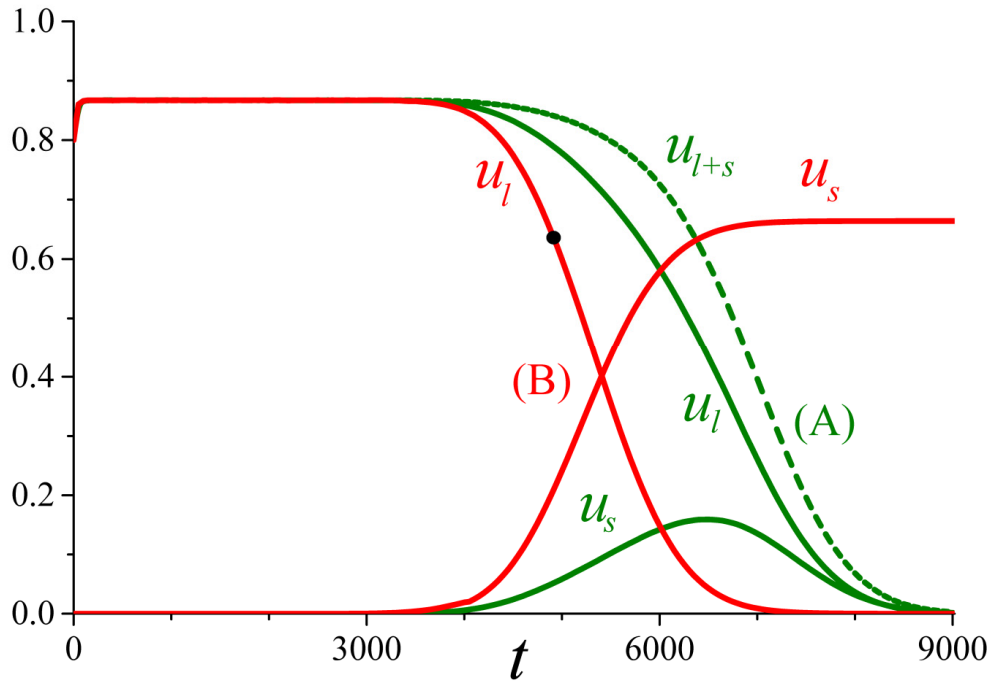

**Figure 3.** (A) The three green curves show one of the calculated results (for 80% initial density) for the total cell density, see Fig, 2, here labeled $u_{l+s}$ (the dashed line), as well as two separate densities (solid lines): healthy live cells, $u_l$, that can still divide (monotonically decreasing, except for very short times), and those cells that can no longer divide, $u_s$ (the peaked curve). (B) As an attempt to include long-living-cell senescence, the two red solid curves show how these separate densities are modified once we make the cells that can no longer divide practically immortal (unchanging on the considered time scales). The density of the healthy live cells, $u_l$, is still monotonically decreasing (except for very short times), whereas that of the senescent cells, $u_s$, is now monotonically increasing. The dot marks the instance of time for which configuration snapshots are shown in Fig. 4.

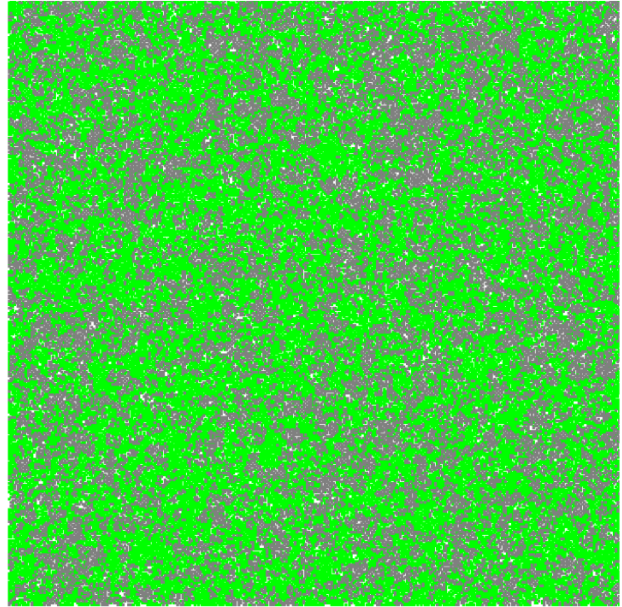
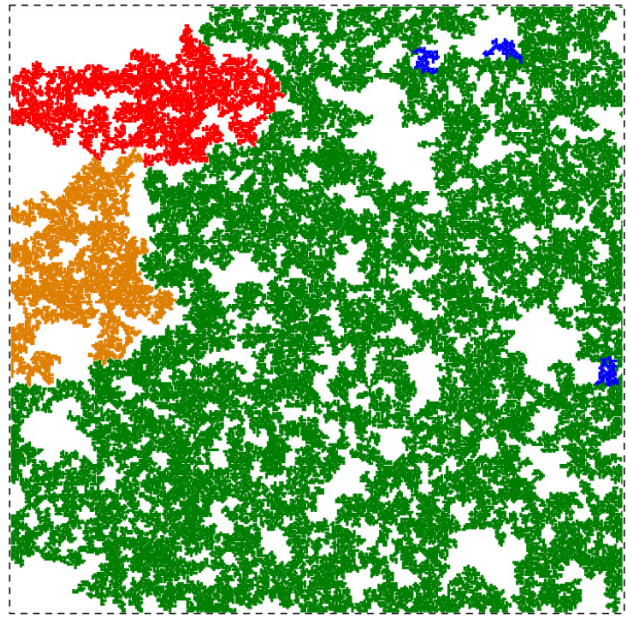

**Figure 4.** For the model with senescence, at the time marked by the solid black dot in Fig. 3, we show cell/cluster snapshots (size 350 × 350 cutouts from a typical 500 × 500 configuration obtained in a random MC run). The top panel shows all the healthy live cells (sites) in bright green, all the senescent cells in gray, and the vacant sites in white. The bottom panel shows only the sites in the three largest (in the count of cells/sites) *connected* healthy live cell clusters, in green, orange, and red, and the sites in the three largest *connected* senescent cell clusters: all in blue.

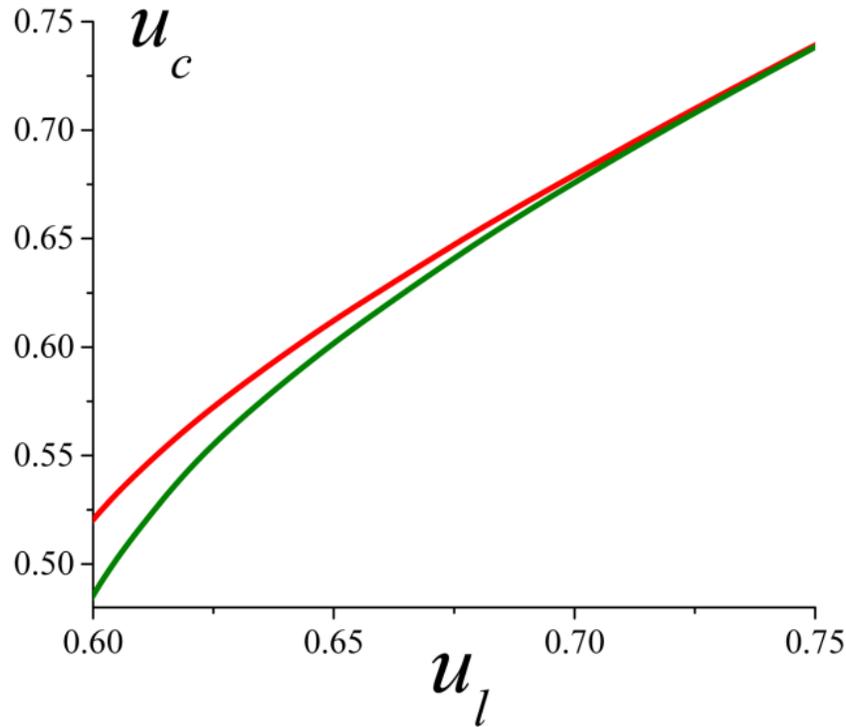

**Figure 5.** This plot shows the density of the healthy live cells that are part of the largest connected cluster of such cells, denoted $u_c$, plotted as a function of the total density of the healthy live cells (that can still divide), $u_l$. After reaching the steady-state plateau, the densities of the healthy live cells, $u_l$, shown for the two considered models in Fig. 3, decrease monotonically for larger times. For the two models: with long-lived senescent cells (the top curve here) and with "senescent" cells dying at the same rate as the regular cells (the bottom curve here), $u_c$ is shown here for the values of $u_l$ once the latter dropped noticeably below the steady-state level (see Fig. 3), but that are still not too small so that the connected live-cell cluster density ($u_c$) is non-negligible for both models. Note that the same $u_l$ values are obtained for different times for the two models; cf. Fig. 3.

For the latter model, with senescence, at the instance of time marked by the solid black dot on the respective $u_l(t)$ curve in Fig 3, we show cell/cluster snapshots in Fig. 4. We note that, while the system is densely populated with cells, the cluster structure of the healthy live cells is already becoming sparse at the selected time, because we are approaching the percolation transition beyond which there is no longer a single system-spanning connected cluster of healthy live sites dominating their morphology. While on average the count of healthy live cells decreases slower with senescence off (supposedly a longer lifespan), there may be tradeoffs in the tissue connectivity (integrity) involved in giving up senescence. This is demonstrated in Fig. 5, which shows the density of those sites which are in the largest connected healthy live cell cluster, $u_c$, as a function of the total density of such cells, $u_l$, for the two cases: cell death with cell division but no long-lived senescence, and with the added senescence as immortality of the cells that no longer divide. We observe that, for the same total healthy live cell densities (actually attained at different times for the two models), the amount of cells that are in the largest *connected* cluster of healthy live cells is in fact increased when senescence is allowed.

## Discussion

### *Potential experiments and model assumptions*

Cells, such as skin cells, are obviously not fully regular, and their shapes are also not symmetrical with respect to their arrangement in the tissue [63]. Assessing the histological preparation of epidermis [71], however, one can argue that lattice-type ordering makes a good approximation for evaluating average quantities, especially considering that we are not interested in long-range fluctuations and correlations. Before discussing this matter further, let us mention a potential experimental setup for verifying percolation model approach and predictions, illustrated in Fig. 6.

In this example, we illustrate how primary human cell dermal fibroblasts can be grown in a cell culture until they reach confluency to closely relate to our 2D model: Fibroblasts have strong contact inhibition properties and stop dividing after forming a dense monolayer of cells [72]. Here the growth was at physiological oxygen (3.5%) in 5% $CO_2$ incubator in 100 mm dishes. Experimental manipulations, such as local physical damage, exposure to UV, or chemotherapeutic drugs can be performed in this setup to collect data, such as growth rate, frequency of senescent and apoptotic cells, to accurately establish various parameters for modeling. In fact, primary cell cultures are generally commonly used in cancer research, and present excellent models to understand and measure basic cell properties, such as the impact of stressors (UV irradiation or mito-toxic drugs) on cell division rate and apoptosis, which will be necessary to estimate and fine-tune the various model parameters.

While it is outside the scope of the present theoretical work, we point out that, specifically for the system depicted in Fig. 6, after the cells formed a dense monolayer they can be stressed by one of the insults described above (physical damage by scratching, UV exposure, and exposure to mito-toxic drugs) or mock handled as the control. Replicates of the cell cultures can then be analyzed at different time points after the treatment.

Another type of experimentally realizable external influence that can be modelled is an introduction of a different type of cells, such as stem cells, following the injury that had induced senescence in the considerable fraction of the existing monolayer. One can create autologous stem cells by reprogramming (induced pluripotency, iPS) of sub-fractions of the fibroblasts by transfection of Yamanaka factors (Oct3/4, Sox2, Klf4, c-Myc) [73]. Additionally, iPS cells will be tagged with GFP protein to track their fate [74]. These cells will be added on top of the existing fibroblast monolayer and participate in the injury repair. Such manipulation can serve as the model of certain stem cell therapies, and understanding of its dynamics by modeling can guide the development of more effective rejuvenation therapies.

Analysis can involve microscopic assessment of cell layer integrity as well as histological assessment of the layer conditions, such as distribution and frequency of the senescent and apoptotic cells. Cell viability can be assessed by propidium iodide exclusion [75], senescent cells can be visualized by betta-gal staining [76], and apoptotic cells can be visualized by terminal deoxynucleotidyl transferase dUTP nick end labeling (TUNEL) [77]. Experiments can then be repeated with cells of early passages (reminiscent of young tissues) and late passages (reminiscent of aging tissues) in order to better understand the changes in cell dynamics that occur with age. In the stem cell experiments, the contribution of externally supplied cells (stem cells) versus original cells (fibroblasts) will be evaluated by means of fluorescent microscopy.

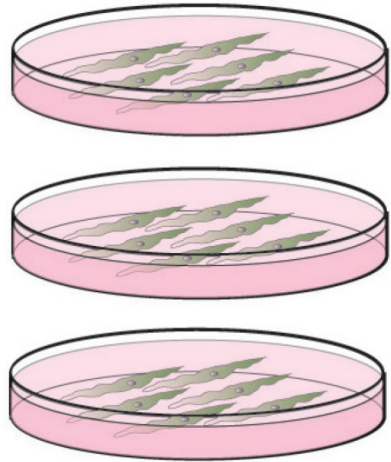

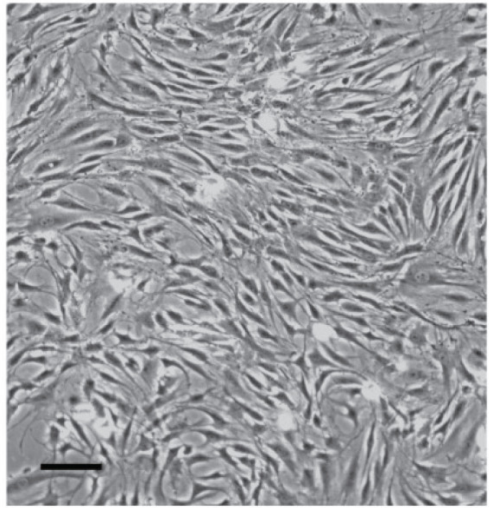

**Figure 6.** A potential experimental setup to determine parameters and test predictions of the percolation aging model. Cells, here human fibroblasts (schematically shown in the top panel) can be grown to confluency. An image of a monolayer of fibroblasts thus grown is shown (in the bottom panel); the bar is 100 μm. The cells can then be damaged or stressed by various mechanisms being modelled, and the progression of the system can be assessed, as described in the text.

We note that, relatively rigid and approximately evenly shaped (so-called "isomeric") objects form random close packings of coordination number approximately 4 in 2D, and 6 in 3D. Therefore, the square (in 2D) and cubic (in 3D) lattices provide a good approximation of the local contact count, if not the random geometry, for such systems. However, even without detailed experiments proposed in the preceding discussion, inspection of Fig. 6 suggests that cells are not evenly shaped and might not be exactly randomly close-packed. Therefore, part of the future experimental work should be devoted to clarifying the applicability of the regular lattice models.

Our model is easily extended to lattice/connectivity-rule structures with varying degrees of randomness and non-uniformity in the nearby site-to-site connectivities, as well as to considering percolation of off-lattice objects [78, 79]. The property of the cell asymmetry can be modelled by assuming cell-cell "interactions" that depend not only on their separation in the count of the lattice spacings, but also on their relative orientation with respect to each other. The primary technical challenges for all such model modifications are that, they require introduction of additional parameters and perhaps MC simulation of increased numerical complexity. Therefore, the results might actually be more difficult to compare to any quantitative or qualitative experimental observations.

*Heterogeneous and external effects, and possible model extensions*

The homogeneous-system modeling could aim at understanding the naturally evolved mechanisms for durable "viability" in the process of aging, while controlling overly proliferate growth of cells (that can lead to cancer) and healing local/limited damage. This is primarily accomplished by cell senescence [80] and cell-cell interactions. However, other effects can also be considered in similar modeling frameworks.

One example is external influence, such as physical damage. From a single cell point of view, some external damage can be rather "large," such as a cut through the skin. Such damage would create a heterogeneity in the cell population, and result in the activation of the body's natural healing response — cell migration to the damaged area. External damage can also be more distributed, such as physical (sunburn) or chemical influences, some of which can result from changes in the body's own conditions in response to external stimulus, and be internal or innate, rather than come from the environment. Some (bio)chemical influences originating externally to the tissues need not be damaging, but could cause accelerated healing and other effects. For example, deletion or pharmaceutical inactivation of α2A/α2C-adrenoceptors in animals greatly accelerates cutaneous wound healing [81] by suppressing the inflammation and accelerating the rate of cell divisions. Modeling by modifying initial conditions, such as for local cuts or sources of inflammation (mosquito bite), and by introducing time-dependent rate changes can then be carried out in the framework of the percolation-model approaches.

A particularly important case for modeling will be the effects of medicines, for example, those for the treatment of cancer. Chemotherapy treatments tend to target fast dividing cells by interfering with DNA synthesis, chromosomal segregation, or growth signals. Most common drugs, such as cyclophosphamide and cisplatin [82], induce DNA damage during cell division, and thus cause apoptosis when the cell attempts to divide. To model presence of such drugs, Rule 1 (cell division) described above, will have an additional probability of both daughter cells undergoing apoptosis as a result of such therapy. Another class of drugs, such as imatinib [83] and erlotinib [84] are designed to suppress activity of growth factors, effectively reducing the rate at which cells divide. To account for the presence of such drugs in our system, Rule 1 will have a greatly reduced probability for cell division. Other interesting compounds with known mechanisms of action can similarly be modelled by modifying Rule 1-8 or introducing additional rules.

Tissue engineering [85-90] frequently involves cell assemblies that can actually be simpler and at least during the processes of their making be also less subjected to external influences than cells in natural tissues. Therefore, they might actually be easier to model within our approach. Furthermore, some tissue-engineering approaches involve [85, 89] scaffolds and other supports that can be incorporated into our model to a large extent as geometric effect (regions of the lattice into which the cells cannot propagate). Consideration of scaffolds, etc., will require modeling of heterogeneous initial conditions and dynamical rules.

## Acknowledgements

We wish to thank Joan Adler and Sergii Domanskyi for useful input and collaboration. The work was in part supported by a seed grant from Cornell Center for Vertebrate Genomics 2014 to SL.

---

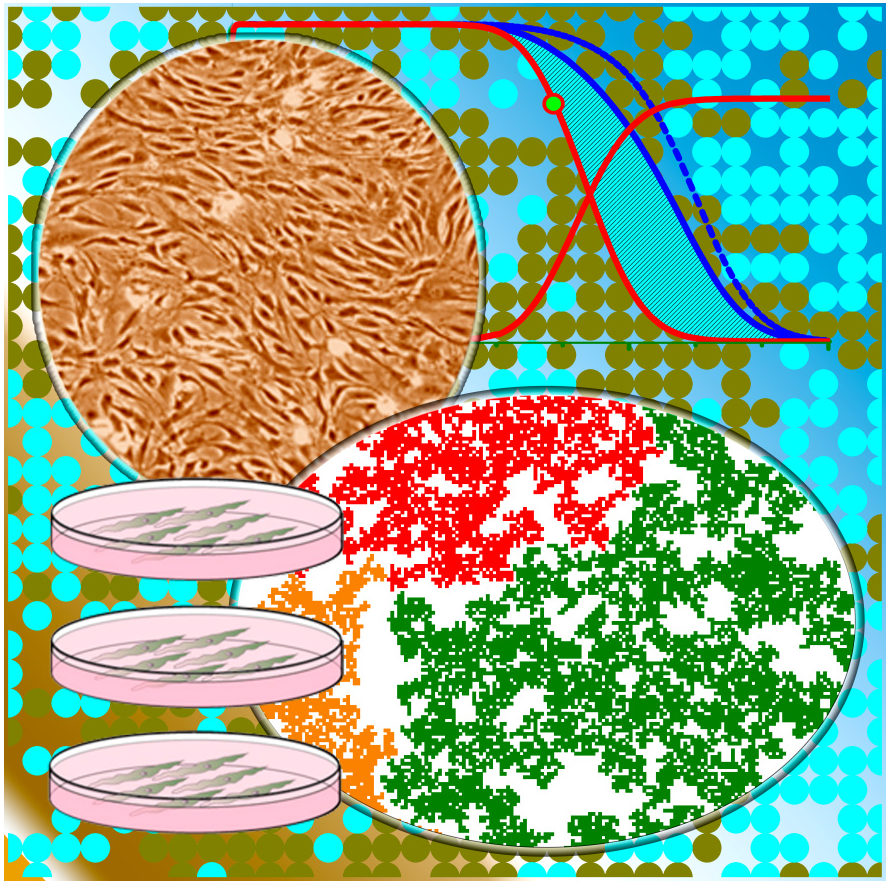

**Punchline:** Percolation modeling is applied to tissue aging, confirming recent experimental finding that inhibition of senescence can lead to extended lifespan.

# References

1. Biala A K, Kirshenbaum L A 2014 The interplay between cell death signaling pathways in the heart. *Trends Cardiovas. Med.* **24** 325-31
2. Burgess A, Rasouli M, Rogers S 2014 Stressing mitosis to death. *Front. Oncol.* **4** 140
3. Mori R, Tanaka K, de Kerckhove M, Okamoto M, Kashiyama K, Tanaka K, Kim S, Kawata T, Komatsu T, Park S, Ikematsu K, Hirano A, Martin P, Shimokawa I 2014 Reduced FOXO1 Expression Accelerates Skin Wound Healing and Attenuates Scarring. *Am. J. Pathol.* **184** 2465-79
4. Campisi J, Robert L 2014 Cell senescence: role in aging and age-related diseases. *Interdiscip. Top. Gerontol.* **39** 45-61
5. Kirkpatrick S 1973 Percolation and Conduction. *Rev. Mod. Phys.* **45** 574-88
6. Stauffer D, Aharony A 1992 *Introduction to percolation theory. 2nd ed.* (London: Taylor & Francis)
7. Dementsov A, Privman V 2007 Percolation modeling of conductance of self-healing composites. *Physica A.* **385** 543-50
8. Dementsov A, Privman V 2008 Three-dimensional percolation modeling of self-healing composites. *Phys. Rev. E* **78** 021104 (6pp)
9. Privman V, Dementsov A, Sokolov I 2007 Modeling of self-healing polymer composites reinforced with nanoporous glass fibers. *J. Comput. Theor. Nanosci.* **4** 190-3
10. Domanskyi S, Privman V 2014 Percolation modeling of self-damaging of composite materials. *Physica A* **405** 1-9
11. Chvapil M, Pfister T, Escalada S, Ludwig J, Peacock E E 1979 Dynamics of the Healing of Skin Wounds in the Horse as Compared with the Rat. *Exp. Mol. Pathol.* **30** 349-59.
12. Debiasio R, Bright G R, Ernst L A, Waggoner A S, Taylor D L 1987 5-Parameter Fluorescence Imaging - Wound-Healing of Living Swiss 3t3 Cells. *J. Cell Biol.* **105** 1613-22
13. Zahm J M, Pierrot D, Chevillard M, Puchelle E 1992 Dynamics of Cell-Movement during the Wound Repair of Human Surface Respiratory Epithelium. *Biorheology.* **29** 459-65
14. Sano H, Ichioka S, Sekiya N 2012 Influence of Oxygen on Wound Healing Dynamics: Assessment in a Novel Wound Mouse Model under a Variable Oxygen Environment. *PLoS ONE* **7** e50212 (7pp)
15. Hwang S W, Kim D H, Tao H, Kim T I, Kim S, Yu K J, Panilaitis B, Jeong J W, Song J K, Omenetto F G, Rogers J A 2013 Materials and Fabrication Processes for Transient and Bioresorbable High-Performance Electronics. *Adv. Funct. Mater.* **23** 4087-93
16. Li R, Cheng H T, Su Y W, Hwang S W, Yin L, Tao H, Brenckle M A, Kim D H, Omenetto F G, Rogers J A, Huang Y G 2013 An Analytical Model of Reactive Diffusion for Transient Electronics. *Adv. Funct. Mater.* **23** 3106-14
17. Toohey K S, Sottos N R, Lewis J A, Moore J S, White S R 2007 Self-healing materials with microvascular networks. *Nature Mater.* **6** 581-5
18. Tee B C K, Wang C, Allen R, Bao Z N 2012 An electrically and mechanically self-healing composite with pressure- and flexion-sensitive properties for electronic skin applications. *Nature Nanotechnol.* **7** 825-32
19. White S R, Sottos N R, Geubelle P H, Moore J S, Kessler M R, Sriram S R, Brown E N, Viswanathan S 2001 Autonomic healing of polymer composites. *Nature* **409** 794-7
20. Cordier P, Tournilhac F, Soulie-Ziakovic C, Leibler L 2008 Self-healing and thermoreversible rubber from supramolecular assembly. *Nature* **451** 977-80

21. Kirk J G, Naik S, Moosbrugger J C, Morrison D J, Volkov D, Sokolov I 2009 Self-Healing Epoxy Composites Based on the Use of Nanoporous Silica Capsules. *Int. J. Fracture* **159** 101-2
22. Kievsky Y, Sokolov I 2005 Self-assembly of uniform nanoporous silica fibers. *IEEE Trans. Nanotechnol*. **4** 490-4
23. Amendola V, Meneghetti M 2009 Self-healing at the nanoscale. *Nanoscale* **1** 74-88
24. Brown E N, Sottos N R, White S R 2002 Fracture testing of a self-healing polymer composite. *Exp. Mech*. **42** 372-9
25. Brown E N, White S R, Sottos N R 2004 Microcapsule induced toughening in a self-healing polymer composite. *J. Mater. Sci*. **39** 1703-10
26. Brown E N, White S R, Sottos N R 2005 Retardation and repair of fatigue cracks in a microcapsule toughened epoxy composite - Part 1: Manual infiltration. *Compos. Sci. Technol*. **65** 2466-73
27. Brown E N, White S R, Sottos N R 2006 Fatigue crack propagation in microcapsule-toughened epoxy. *J. Mater. Sci*. **41** 6266-73
28. Kessler M R, Sottos N R, White S R 2003 Self-healing structural composite materials. *Compos. Part A* **34** 743-53
29. Tan P S, Zhang M Q, Bhattacharyya D 2009 Processing and Performance of Self-Healing Materials. *IOP Conf. Ser.: Mater. Sci*. **4** 012017 (5pp)
30. Pang J W C, Bond I P 2005 A hollow fibre reinforced polymer composite encompassing self-healing and enhanced damage visibility. *Compos. Sci. Technol*. **65** 1791-9
31. Piermattei A, Karthikeyan S, Sijbesma R P 2009 Activating catalysts with mechanical force. *Nature Chem*. **1** 133-7.
32. Shchukin D G, Mohwald H 2007 Self-repairing coatings containing active nanoreservoirs. *Small*. **3** 926-43.
33. Williams G, Trask R, Bond I 2007 A self-healing carbon fibre reinforced polymer for aerospace applications. *Compos. Part A* **38** 1525-32
34. Balazs A C 2007 Modelling self-healing materials. *Mater. Today* **10** 18-23
35. Kolmakov G V, Matyjaszewski K, Balazs A C 2009 Harnessing Labile Bonds between Nanogel Particles to Create Self-Healing Materials. *ACS Nano*. **3** 885-92
36. Barbero E J, Greco F, Lonetti P 2005 Continuum damage-healing mechanics with application to self-healing composites. *Int. J. Damage Mech*. **14** 51-81
37. Murphy E B, Wudl F 2010 The world of smart healable materials. *Prog Polym Sci*. **35** 223-51
38. Wool R P 2008 Self-healing materials: a review. *Soft Matter* **4** 400-18
39. Fedrizzi L, Fürbeth W, Montemor F, European Federation of Corrosion, Institute of Materials Minerals and Mining, Knovel (Firm) 2011 *Self-healing properties of new surface treatments* (Leeds: Published for the European Federation of Corrosion by Maney Pub. on behalf of the Institute of Materials, Minerals & Mining)
40. Zhang M Q, Rong M Z 2011 *Self-healing polymers and polymer composites* (Hoboken, New Jersey: Wiley)
41. Thostenson E T, Chou T W 2006 Carbon nanotube networks: Sensing of distributed strain and damage for life prediction and self healing. *Adv. Mater*. **18** 2837-41
42. Bregg R K 2006 *Frontal polymer research* (New York: Nova Science Publishers)
43. LaPak K M, Burd C E 2014 The Molecular Balancing Act of p16(INK4a) in Cancer and Aging. *Mol. Cancer Res*. **12** 167-83

44. Blagoev K B 2011 Organ aging and susceptibility to cancer may be related to the geometry of the stem cell niche. *Proc. Nat. Acad. Sci. USA* **108** 19216-21
45. Antal T, Blagoev K B, Trugman S A, Redner S 2007 Aging and immortality in a cell proliferation model. *J. Theor. Biol*. **248** 411-7
46. Sahimi M 1994 *Applications of percolation theory. London* (Bristol: Taylor & Francis)
47. Macheras P, Iliadis A 2006 *Modeling in Biopharmaceutics, Pharmacokinetics, and Pharmacodynamics. Homogeneous and Heterogeneous Approaches* (New York, NY: Springer, Inc.)
48. Baker D J, Wijshake T, Tchkonia T, LeBrasseur N K, Childs B G, van de Sluis B, Kirkland J L, van Deursen J M 2011 Clearance of p16(Ink4a)-positive senescent cells delays ageing-associated disorders. *Nature* **479** 232-6
49. Du W W, Yang W, Fang L, Xuan J, Li H, Khorshidi A, Gupta S, Li X, Yang B B 2014 miR-17 extends mouse lifespan by inhibiting senescence signaling mediated by MKP7. *Cell Death Dis*. **5** e1355 (14pp)
50. Jacobs D J, Thorpe M F 1996 Generic rigidity percolation in two dimensions. *Phys. Rev. E* **53** 3682-93
51. Feng S, Sen P N 1984 Percolation on Elastic Networks - New Exponent and Threshold. *Phys. Rev. Lett*. **52** 216-9
52. Wool R P 2005 Rigidity percolation model of polymer fracture. *J. Polym. Sci. Pol. Phys*. **43** 168-83
53. Thorpe M F, Jacobs D J, Chubynsky M V, Phillips J C 2000 Self-organization in network glasses. *J. Non-Cryst. Solids* **266** 859-66
54. Haynes B, Coles M, Azzi D 2008 A self-healing mobile wireless sensor network using predictive reasoning. *Sensor Rev*. **28** 326-33
55. Qu Y, Georgakopoulos S 2012 An Average Distance Based Self-Relocation and Self-Healing Algorithm for Mobile Sensor Networks. *Wireless Sensor Network* **4** 257-63
56. Chowdhury A R, Tripathy S, Nandi S 2007 Securing wireless sensor networks against spurious injections. *2007 2nd International Conference on Communication Systems Software & Middleware* **1 & 2** 287-91
57. Ovadya Y, Krizhanovsky V 2014 Senescent cells: SASPected drivers of age-related pathologies. *Biogerontology* **15** 627-42
58. Holliday R 2014 The Commitment of Human Cells to Senescence. *Interdisc. Top. Gerontol*. **39** 1-7
59. Goyns M H, Lavery W L 2000 Telomerase and mammalian ageing: a critical appraisal. *Mech. Ageing Devel*. **114** 69-77.
60. Libert S, Chao Y F, Chu X W, Pletcher S D 2006 Trade-offs between longevity and pathogen resistance in Drosophila melanogaster are mediated by NF kappa B signaling. *Aging Cell* **5** 533-43
61. Zhang M J, Zhou J, Wang L, Li B, Guo J W, Guan X, Han Q J, Zhang H J 2014 Caffeic Acid Reduces Cutaneous Tumor Necrosis Factor Alpha (TNF-alpha), IL-6 and IL-1 beta Levels and Ameliorates Skin Edema in Acute and Chronic Model of Cutaneous Inflammation in Mice. *Biol. Pharm. Bull*. **37** 347-54
62. Montgomery C P, Daniels M D, Zhao F, Spellberg B, Chong A S, Daum R S 2013 Local Inflammation Exacerbates the Severity of Staphylococcus aureus Skin Infection. *PLoS ONE* **8** e69508 (6pp)

63. Rygiel T P, Mertens A E, Strumane K, van der Kammen R, Collard J G 2008 The Rac activator Tiam1 prevents keratinocyte apoptosis by controlling ROS-mediated ERK phosphorylation. *J. Cell Sci*. **121** 1183-92
64. Sinkovics J G 1991 Programmed Cell-Death (Apoptosis) - Its Virological and Immunological Connections (a Review). *Acta Microbiol. Hung*. **38** 321-34
65. Sangiuliano B, Perez N M, Moreira D F, Belizario J E 2014 Cell Death-Associated Molecular-Pattern Molecules: Inflammatory Signaling and Control. *Mediat. Inflamm*. **2014** 821043 (14pp)
66. Allan D J, Harmon B V 1986 The morphologic categorization of cell death induced by mild hyperthermia and comparison with death induced by ionizing radiation and cytotoxic drugs. *Scan. Electron Microsc*. **3** 1121-33
67. Yoshihisa Y, Rehman M U, Shimizu T 2014 Astaxanthin, a xanthophyll carotenoid, inhibits ultraviolet-induced apoptosis in keratinocytes. *Exp. Dermatol*. **23** 178-83
68. Salminen A, Ojala J, Kaarniranta K 2011 Apoptosis and aging: increased resistance to apoptosis enhances the aging process. *Cell. Mol. Life Sci*. **68** 1021-31
69. Vjetrovic J, Shankaranarayanan P, Mendoza-Parra M A, Gronemeyer H 2014 Senescence-secreted factors activate Myc and sensitize pretransformed cells to TRAIL-induced apoptosis. *Aging Cell* **13** 487-96
70. Rubin H 2002 The disparity between human cell senescence in vitro and lifelong replication in vivo. *Nature Biotechnol*. **20** 675-81
71. Symonette C J, Mann A K, Tan X C, Tolg C, Ma J, Perera F, Yazdani A, Turley E A 2014 Hyaluronan-Phosphatidylethanolamine Polymers Form Pericellular Coats on Keratinocytes and Promote Basal Keratinocyte Proliferation. *Biomed. Res. Int*. **2014** 727459 (14pp)
72. Boss J M N, Dessy C 1969 Behaviour of Cells Arrested by Contact Inhibition. *J Physiol-London*. **203** P28
73. Yamanaka S 2009 A Fresh Look at iPS Cells. *Cell* **137** 13-7
74. Kain S R, Kitts P 1997 Expression and detection of green fluorescent protein (GFP). *Methods Mol. Biol*. **63** 305-24
75. Dengler W A, Schulte J, Berger D P, Mertelsmann R, Fiebig H H 1995 Development of a Propidium Iodide Fluorescence Assay for Proliferation and Cytotoxicity Assays. *Anticancer Drug*s **6** 522-32.
76. Debacq-Chainiaux F, Erusalimsky J D, Campisi J, Toussaint O 2009 Protocols to detect senescence-associated beta-galactosidase (SA-beta gal) activity, a biomarker of senescent cells in culture and in vivo. *Nature Protoc*. **4** 1798-806
77. Gavrieli Y, Sherman Y, Ben-Sasson S A 1992 Identification of programmed cell death in situ via specific labeling of nuclear DNA fragmentation. *J. Cell Biol*. **119** 493-501
78. Balberg I, Wagner N, Goldstein Y, Weisz S Z 1990 Tunneling and Percolation Behavior in Granular Metals. *Mater. Res. Soc. Symp. P* **195** 233-8
79. Drory A, Berkowitz B, Parisi G, Balberg I 1997 Theory of continuum percolation .3. Low-density expansion. *Phys. Rev. E* **56** 1379-95
80. Serrano M, Lin A W, McCurrach M E, Beach D, Lowe S W 1997 Oncogenic ras provokes premature cell senescence associated with accumulation of p53 and p16(INK4a). *Cell* **88** 593-602
81. Romana-Souza B, Nascimento A P, Brum P C, Monte-Alto-Costa A 2014 Deletion of the alpha 2A/alpha 2C-adrenoceptors accelerates cutaneous wound healing in mice. *Int. J. Exp. Pathol*. **95** 330-41

82. Antunovic M, Kriznik B, Ulukaya E, Yilmaz V T, Mihalic K C, Madunic J, Marijanovic I 2015 Cytotoxic activity of novel palladium-based compounds on leukemia cell lines. *Anticancer Drugs* **26** 180-6
83. Wang W L, Healy M E, Sattler M, Verma S, Lin J, Maulik G, Stiles C D, Griffin J D, Johnson B E, Salgia R 2000 Growth inhibition and modulation of kinase pathways of small cell lung cancer cell lines by the novel tyrosine kinase inhibitor STI 571. *Oncogene* **19** 3521-8
84. Bulgaru A M, Mani S, Goel S, Perez-Soler R 2003 Erlotinib (Tarceva): a promising drug targeting epidermal growth factor receptor tyrosine kinase. *Expert Rev. Anticancer Ther*. **3** 269-79
85. Dvir T, Timko B P, Kohane D S, Langer R 2011 Nanotechnological strategies for engineering complex tissues. *Nature Nanotechnol*. **6** 13-22
86. Murphy S V, Atala A 2014 3D bioprinting of tissues and organs. *Nature Biotechnol*. **32** 773-85
87. Marga F, Jakab K, Khatiwala C, Shepherd B, Dorfman S, Hubbard B, Colbert S, Forgacs G 2012 Toward engineering functional organ modules by additive manufacturing. *Biofabrication* **4** 022001 (12pp)
88. Naderi H, Matin M M, Bahrami A R 2011 Review paper: Critical Issues in Tissue Engineering: Biomaterials, Cell Sources, Angiogenesis, and Drug Delivery Systems. *J. Biomater. Appl*. **26** 383-417
89. Tran R T, Thevenot P, Zhang Y, Gyawali D, Tang L P, Yang J 2010 Scaffold Sheet Design Strategy for Soft Tissue Engineering. *Mater*. **3** 1375-89
90. Zheng W, Jiang X 2014 Precise manipulation of cell behaviors on surfaces for construction of tissue/organs. *Coll. Surf. B* **124** 97-110